
\magnification=1200
\baselineskip=15pt
\hsize=6truein
\def\CALT#1{\hbox to\hsize{\tenpoint \baselineskip=12pt
	\hfil\vtop{\hbox{\strut CALT-68-#1}
	\hbox{\strut DOE RESEARCH AND}
	\hbox{\strut DEVELOPMENT REPORT}}}}

\def\ABSTRACT#1{\vskip .5in \vfil \centerline{\twelvepoint \bf Abstract}
	#1 \vfil}
\def\ENDTITLEPAGE{\vfil\eject\pageno=1}

\def\sqr#1#2{{\vcenter{\hrule height.#2pt
      \hbox{\vrule width.#2pt height#1pt \kern#1pt
        \vrule width.#2pt}
      \hrule height.#2pt}}}

\def\section#1#2{
\noindent\hbox{\hbox{\bf #1}\hskip 10pt\vtop{\hsize=5in
\baselineskip=12pt \noindent \bf #2 \hfil}\hfil}
\medskip}

\def\underwig#1{	
	\setbox0=\hbox{\rm \strut}
	\hbox to 0pt{$#1$\hss} \lower \ht0 \hbox{\rm \char'176}}

\def\bunderwig#1{	
	\setbox0=\hbox{\rm \strut}
	\hbox to 1.5pt{$#1$\hss} \lower 12.8pt
	 \hbox{\seventeenrm \char'176}\hbox to 2pt{\hfil}}

\def\MEMO#1#2#3#4#5{
\frontpagetrue
\centerline{\tencp INTEROFFICE MEMORANDUM}
\smallskip
\centerline{\bf CALIFORNIA INSTITUTE OF TECHNOLOGY}
\bigskip
\vtop{\tenpoint
\hbox to\hsize{\strut \hbox to .75in{\caps to:\hfil}\hbox to 3.8in{#1\hfil}
\quad\the\date\hfil}
\hbox to\hsize{\strut \hbox to.75in{\caps from:\hfil}\hbox to 3.5in{#2\hfil}
\hbox{{\caps ext-}#3\qquad{\caps m.c.\quad}#4}\hfil}
\hbox{\hbox to.75in{\caps subject:\hfil}\vtop{\parindent=0pt
\hsize=3.5in #5\hfil}}
\hbox{\strut\hfil}}}
\tolerance=10000
\hfuzz=5pt
\rightline{UR-1338}
\rightline{ER40685-787}
\rightline {IP/BBSR /94-3}
\rightline {January, 1994}
\centerline  {\bf {Duality of the Superstring in Superspace}}
\bigskip

\centerline{Ashok Das}
\centerline{Department of Physics and Astronomy}
\centerline{University of Rochester}
\centerline{Rochester, NY 14627}
\medskip
\centerline{and}
\medskip
\centerline{Jnanadeva  Maharana}
\centerline{Institute of Physics}
\centerline{Bhubaneswar, India}

\ABSTRACT{The evolution of a closed NSR string is considered
in the background of constant graviton and antisymmetric fields. The
$\sigma$-model action is written in a manifestly supersymmetric form
in terms of superfields. The first order formalism adopted for the
closed bosonic string is generalised to implement duality
transformations and the constant dual backgrounds are obtained for
the dual theory. We recover the $G \rightarrow G^{-1}$ duality for
the case when antisymmetric tensor field is set to zero. Next, the
case when the backgrounds depend on one superfield, is also analysed. This
scenario is similar to the cosmological case envisaged for the
bosonic string. The explicit form of the duality transformation is
given for this case. }

\ENDTITLEPAGE

\eject
Recently, considerable attention has been focused on the study of various
target space symmetries of string theory. The duality$^1$ symmetry is
one of the most well known symmetries. The essential features of
duality (the so called $R$-duality) were discussed almost a decade ago.
  One of the consequences of $R$-duality is that the dynamics of a string on a
 circle
of radius $R$ is equivalent to that of another string on a circle of
radius
 ${1\over R}$  (in suitable units).  The idea of duality has also
been used in a wider context$^{2,3}$ and it is conjectured that this symmetry
might be maintained to all orders in string perturbation theory.$^4$
Furthermore, duality has been applied to study string cosmology$^{5-7}$
and used to obtain new black hole solutions.$^7$

Recently the concept of scale factor duality$^6$ (SFD) has been
introduced as a symmetry group of classical string equations of motion,
derived from a low energy string effective action.  One of the salient
features of SFD is that it does not require compactification of the target
space.  Moreover, this transformation relates different time-dependent
background configurations of string theory.  There is an intimate
connection between Narain's$^8$ construction of inequivalent static
compactification and the $O(d,d)$ transformations on background fields, which
rotate time-dependent backgrounds (solutions of equations of motion)
into other ones which are not necessarily equivalent.  Subsequently,
several cosmological$^9$ and black hole solutions$^{10}$ have been
obtained through the implementation of $O(d,d)$ transformations and
their generalizations.  It has been observed that the roles of the
canonical momentum $P$ and $X'$ are interchanged under duality (this
amounts to an interchange of winding number and momentum zero
modes).Moreover, when we consider evolution of the string in
constant background fields, it is necessary to transform the
backgrounds appropriately along with the interchange of
$P\leftrightarrow X'$ so that Hamiltonian remains invariant under duality.

We note that for constant background field configurations, when we
consider a closed string with toroidal compactification of
$d-$coordinates, the compact coordinates satisfy periodicity
conditions. It follows from the single-valuedness of the wave function
that the zero modes of the canonical momenta corresponding to these
compactified coordinates take integer values. As mentioned above, $P$
and $X'$ are interchanged under duality transformation and
consequently winding number and zero modes of these momenta get
interchanged. Veneziano$^6$ and his collaborators$^9$ have considered
the case when the backgrounds depend only on time, and have shown
that the tree level string effective action is invariant under global
$O(d,d)$ trnsformations, $d$ being number of spatial dimensions. It
has been noted in refs. 6 and 9 that, unlike the case of Narain
compactification, it is not necessary that the spatial dimensions be
compactified to obtain $O(d,d)$ invariance of the string effective
action.

The duality and noncompact O(d,d) symmetries of string theory have
been  investigated from the worldsheet point of view recently $^{11,12}$.
It was found useful to adopt a first order formalism and then
introduce dual string coordinates corresponding to the string
coordinates and obtain the corresponding Lagrangian involving these
coordinates. The evolution equations for the dual variables were
derived. In the recent past target space symmetry properties, such as
duality and $O(d,d)$ transformations, have also been studied by several
authors$^{13-17}$.

The purpose of this note is to generalize this formalism to
superstrings. First, the evolution of the string is considered in the
constant backgrounds of graviton and antisymmetric tensor fields
denoted by $G$ and $B$ respectively (we shall use greek alphabets
$\mu, \nu$ etc as spacetime indices when we intend to specify the
components explicitly). The worldsheet action is expressed in terms
of  superfields in a manifestly supersymmetric invariant form.
Subsequently, a passage to the first order formalism is achieved by
introducing a spinor superfield whereas the original action involves
only scalar superfields (whose components are identified with
bosonic and fermionic coordinates of the string). The equations of
motion for the fermionic and scalar superfield can be used to derive
the evolution equations for the string on the worldsheet for constant
background fields. Next, the dual superfield is introduced and the
corresponding first order Lagrangian is obtained. The equation of
motion for the dual coordinate reveals that the corresponding
background fields $\tilde G$ and $\tilde B$ are the metric and the
antisymmetric tensor fields associated with the dual coordinates.
Next, we consider the situation where the backgrounds depend only on
 one superfield instead of being constant. In this case, the
procedure of ref. 11 can be utilised to derive the evolution equations
and implement the duality transformations to obtain the dual
coordinates. The superspace action has the form

$$ I = - {1\over 2} \int d^2\sigma d^2\theta \bar D{\phi^{\mu}} (
G_{\mu\nu}(\phi) - \gamma_5 B_{\mu\nu}(\phi) ) D\phi{^\nu} \eqno  (1) $$

\noindent where the integral $d^2\sigma$ is over the worldsheet
coordinates and we have already gone over to the orthonormal gauge.
The Grassmann variables, $\theta$, represent the fermionic coordinates of
the superspace.
Here $\phi{^\mu}$ is a real scalar superfield which has the form

$$ \phi^{\mu} = X{^\mu} +  \bar \theta \psi{^\mu} + {1\over 2} {\bar \theta}
\theta F{^\mu} \eqno (2) $$

\noindent where $X^{\mu}$ and $\psi^{\mu}$ correspond to the bosonic and
the fermionic string coordinates respectively and we have
suppressed the spinor indices for simplicity. The convention of the
gamma matrices are as follows:

$$\gamma^0 = \pmatrix{0 &1\cr
\noalign{\vskip 5pt}%
1 &0\cr} \qquad
 \gamma^1 = \pmatrix{0 &-1\cr
\noalign{\vskip 5pt}%
1 &0\cr} \qquad
 \gamma_5 =  \gamma^0 \gamma^1 =
\pmatrix{1 &0\cr
\noalign{\vskip 5pt}%
0 &-1\cr} \eqno (3)$$

\noindent The covariant derivatives in the superspace are defined to be

$$D_\alpha = {{\partial \over {\partial  \bar \theta_\alpha}}} - i (
\gamma^ a \theta)_\alpha \partial_a
\qquad  \bar D_\alpha = - {\partial \over \partial \theta_\alpha}
+ i ( \bar \theta \gamma^ a)_\alpha \partial_a \eqno (4) $$

\noindent Here $\bar \theta = \theta^\dagger \gamma^0$ and
$\partial_a$ with $a = 0,1$ correspond to derivatives with
respect to the world sheet coordinates $\tau$ and $\sigma$.
Notice that the backgrounds, in general, depend on the
superfields and the conventional Lagrangian can be obtained by expanding the
backgrounds in terms of the
 components of the  superfields and then carrying out the
integrations over the real two component Majorana spinor $\theta$.
The equations of motion are

$$\bar D \left( G_{\mu \nu} (\phi) - \gamma_5 B_{\mu \nu}
(\phi) \right) D \phi^\nu = 0 \eqno (5)$$

\noindent In the special case when backgrounds are independent of the
superfields, it is quite evident that the equations of motion take
the form of a current conservation  since the Lagrangian only depends
the derivatives of the fields.

$$ \bar D ( G_{\mu\nu} - \gamma_5 B_{\mu\nu} ) D\phi^{\nu} = 0
\eqno (6) $$

\noindent We may recall that this equation is similar to an equation
$ \partial_a j^a = 0 $ obtained for the bosonic string,
where the derivative is with respect to the worldsheet coordinates
and $j^a$ is the derivative of the corresponding Lagrangian
with respect to $\partial_a X^{\mu}$.

In order to go over to the first order formalism, we introduce a new
Lagrangian  with an auxiliary  Majorana spinor superfield
$ \Psi^{\mu}_{\alpha}$, where $\mu$ and $\alpha$ are the target
space index and the spinorial index respectively.

$$ L_1 = {1\over 2} \bar \Psi^\mu ( G_{\mu\nu} - \gamma_5
B_{\mu\nu} ) \Psi^{\nu} - \bar \Psi^{\mu} ( G_{\mu\nu} - \gamma_5
B_{\mu\nu} ) D \phi^{\nu} \eqno (7) $$

Notice that there is no derivative term of $\Psi^{\mu}$ in the
Lagrangian and consequently, the equation of motion for $\Psi^\mu$ is
merely a constraint equation. The two equations of motion are

$$ {\delta L_1 \over {\delta \bar \Psi^{\mu}}} = ( G_{\mu\nu} -
\gamma_5 B_{\mu\nu} ) ( \Psi^{\nu} - D \phi^{\nu} ) = 0 \eqno (8)$$

\noindent and

$$ {\delta L_1 \over {\delta \phi^{\mu}}} = \bar D ( G_{\mu\nu} -
\gamma_5 B_{\mu\nu} ) \Psi^{\nu} = 0  \eqno (9)$$

It follows from Eq. (8) that $ \Psi^{\mu} = D\phi^{\mu} $ and by
substituting this in Eq. (9) we recover the evolution equation (5) for the
superfield $\phi^{\mu}$.

Now we consider another Lagrangian, $L_2$ and introduce the dual
superfield $\tilde \phi_{\mu}$ with the auxiliary field   denoted
by $\tilde{\Psi}^{\mu}$. Thus

$$ L_2 = {1\over 2} \bar{\tilde{\Psi}}^{\mu} ( G_{\mu\nu} -
\gamma_5 B_{\mu\nu}
)\tilde{\Psi}^{\nu} - \bar{\tilde{\Psi}}^{\mu} D \tilde \phi_\mu \eqno (10)$$

\noindent The equations of motion are

$$ {\delta L_2 \over {\delta \bar{\tilde{\Psi}}^{\mu}}} = ( G_{\mu\nu} -
\gamma_5 B_{\mu\nu} ) \tilde{\Psi}^{\nu} - D \tilde \phi_\mu = 0 \eqno (11) $$

\noindent and

$$ {\delta L_2 \over {\delta \tilde \phi_\mu}} = \bar D \tilde{\Psi}^{\mu} =
0 \eqno (12) $$

We can express $\tilde{\Psi}^{\mu}$ in terms of $\tilde \phi_\mu$ from
Eq. (11)
and then substitute in equations (12) in order to get the equation of
motion for the dual superfield. Let us define

$$ \tilde G = ( B^{-1} G - G^{-1} B )^{-1} B^{-1} \eqno (13) $$

\noindent and

$$ \tilde B = -( B^{-1} G - G^{-1}B )^{-1} G^{-1} \eqno (14)$$

\noindent Notice that both $\tilde G$ and $\tilde B $ are defined
such that they appear with upper indices whereas $G$ and $B$ are
defined with lower indices. And it is easy to check that

$$ (\tilde G - \gamma_5  \tilde B ) ( G - \gamma_5 B ) = \bf 1 \eqno (15) $$

\noindent It is to be interpreted that $\tilde G$ is the metric (is
symmetric) and $\tilde B$ is the corresponding antisymmetric tensor
for the dual theory. Now the equation of motion for the dual
superfield takes the form (supressing spacetime as well as the
spinorial indices)

$$ \bar D (\tilde G - \gamma_5 \tilde B ) D \tilde \phi = 0 \eqno (16)$$

We note that the equations of motion involving the dual variables have the
same form as those for the original superfields. However, now the metric
and the antisymmetric tensor fields are to be replaced by $\tilde G$ and
$\tilde B$ respectively. In order to see the connection with the duality
transformation, associated with  the bosonic string, we set $ B = 0 $ in
the above equation and find that $ G \rightarrow G^{-1} $ under the duality
 transformation as was the case for the bosonic string.

  Next, we take up the case when the background depends
 on only one of the superfields, say $\phi^0$. This situation is analogous
 to the cosmological case where the backgrounds are taken to be
 time dependent only. Here it is easy to see that the equation of motion
 for the $\phi^0$ field involves the variation of the action with
 respect to $\phi^0$ as well as $D \phi^0$ since backgrounds depend
 on this coordinate. The action can be written in a slightly simplified
 manner if we recognise that $ G_{\mu \nu}$ and $ B_{\mu \nu}$ transform
as second rank tensors under general coordinate transformations
and there is an Abelian gauge transformation associated with
$B_{\mu \nu}$ wheru
the gauge function is a super field with a
target space vector index.
For the case at hand where $G$ and $B$ depend on only one
superfield, it is convenient
 to bring $G_{\mu\nu}$ and
$B_{\mu\nu}$ to the following special
form by implementing a
general coordinate transformation and an Abelian gauge
transformation respectively.

$$ G_{\mu\nu} = \left (\matrix { -1 & 0 \cr  0 &G_{ij} (\phi^0)
\cr } \right) \eqno (17a) $$

$$ B_{\mu\nu} = \left (\matrix { 0 & 0 \cr 0 & B_{ij}(\phi^0)
\cr} \right) \eqno (17b) $$

\noindent Here $i,j = 1,\cdots, d$ are indices of the spatial
coordinates. Now the Lagrangian $ \bar {L}$ can be reexpressed
as

$$ \bar{L} = {1 \over 2} \bar{ D} \phi^0 D \phi^0 - { 1 \over
2} \bar{D} \phi^{i} D \phi^{j} G_{ij} + {1 \over 2} \bar {D}
\phi^{i} \gamma_5 B_{ij} D \phi^{j} + L_D \eqno (18) $$

Since $G_{ij}$ and $B_{ij}$ depend only on the superfield
$\phi^0$, the evolution equation for the space component
superfields, $\phi^{i}$, takes the form of Eq. (5) where
the space time indices $\mu,\nu$ are replaced by the spatial
indices $i,j$ running from $1$ to $d$. Here $L_D$ contains the
dilaton coupling to the string. In what follows, we shall
concentrate our discussions only on the first three terms of
Eq. (18).

The equation of motion of the superfields $\phi^{i}$
can be obtained from Eq. (18) and takes  the form

$$ {{\delta \bar {L}} \over {\delta \phi^{i}}} = \bar{ D} (
G_{ij} (\phi^0) - \gamma_5 B_{ij}(\phi^0)) D \phi^{j} = 0 \eqno
(19) $$

This equation follows from Eq. (5) if we restrict to the spatial
indices in the present case. The $\phi^0$ equation is more
complicated and would involve functionally differentiating
$G_{ij}(\phi^0)$. As before, it is straight
forward to introduce a first order Lagrangian involving the
superfields $\phi^{i}$ and a Majorana spinor superfield
$\Psi^{i}_{\alpha}$. The construction of the dual Lagrangian
will go through in a completely parallel manner to the case of
constant backgrounds. The metric and the antisymmetric tensor
superfields associated with the dual Lagrangian will also depend
only on the super field $\phi^0$ and will satisfy

$$ ( \tilde {G}(\phi^0) - \gamma_5 \tilde {B}(\phi^0))^{ij} (
G(\phi^0) - \gamma_5 B(\phi^0))_{jk} = \delta^{i}_{k} \eqno (20)
$$

\noindent The equations of motion for the dual superfields
$\tilde {\phi}^{i}$ will be of the same form as the equations
for the original superfields as has been emphasised in Eq. (16).
The component field eu
ations can be easily obtained from
Eq. (19) through the use of the decomposition in Eq. (2). In the
present case, they take forms

$$ \partial_{a} ( \eta^{ab} G_{ij} + \epsilon^{ab} B_{ij}
)\partial_{b} X^{j} = 0 $$

$$ ( G_{ij} + \gamma_5 B_{ij}) \gamma^{a} \partial_{a} \psi^{j}
= 0  $$

$$ F^{i} = 0 \eqno (21) $$

The component equations for the dual superfields will have
identical forms with the metric, $G_{ij}$, and the antisymmetric
tensor field, $B_{ij}$, replaced by their duals respectively.

Now we proceed to compute the super current in the case of constant
background fields. The case where the backgrounds depend only on
one superfield will be completely parallel. We recall that
Eq. (8) can be used to eliminate the fermionic superfield,
$\Psi^{\mu}$ in terms of $\phi^{\mu}$ and $L_1$ can be expressed
in terms of $\phi^{\mu}$ alone. The super current can be
computed in a straight forward manner using the definition of
the supersymmetry generators in superspace and it has the
following form in terms of component fields

$$ J^{a} = \partial_{b} X^{\mu} \gamma^{b} \gamma^{a} \psi^{\nu}
G_{\mu\nu} \eqno (22) $$

The supercharge densities can be written in the chiral basis as

$$ J^{\pm} = \partial_{\pm} X^{\mu} \psi^{\nu}_{\pm} G_{\mu\nu}
\eqno (23) $$

\noindent where $\psi^{\mu}_{\pm} = { 1 \over 2} ( 1 \mp
\gamma_{5} ) \psi^{\mu} $, $ J^{\pm} = { 1\over 2} ( J^0 \pm J^1
) $ and $ \partial_{\pm} = \partial_0 \pm \partial_1 $.  Notice
that when we consider the case where $ G_{\mu\nu} = \eta_{\mu\nu}
$ and $ B_{\mu\nu} = 0 $, under the duality transformation, $
\partial_{\pm} X^{\mu} \rightarrow {\pm} \partial_{\pm} X^{\mu}
$ and we expect that $\psi^{\mu}_{\pm} \rightarrow {\pm}
\psi^{\mu}_{\pm}$ if the corresponding supercharges are to
remain invariant. For the case at hand, the supercharges
corresponding to the dual theory described by $\tilde
{\phi}^{\mu} $ and $\tilde {\Psi}^{\mu}$ in Eq. (10) can also be
obtained in a similar manner. We can eliminate the spinor superfield
, $\tilde { \Psi}^{\mu}$, from Eq. (11) to write the theory
completely in terms of the superfield $\tilde {\phi}^{\mu}$.
The super current in this case has the form

$$ { \tilde {J}^{a} } = \partial_b \tilde {X}^{\mu} \gamma^b
\gamma^a \tilde {\psi}^{\nu} \tilde {G}_{\mu\nu} \eqno (24) $$

\noindent where $\tilde G $ is given in Eq. (13) and $\tilde X$,
$\tilde \psi$ are the scalar and the spinor components of the
dual superfield $\tilde \phi$.

In this letter, we have investigated the question of duality in
superspace formulation of the superstring. The two cases
considered correspond to constant backgrounds as well as the
case where the backgrounds depend only on one of the superfields.
The question of duality transformation is particularly tricky in
superspace. Note that in the case of a bosonic string, in
constant background, the evolution equation has the form of a
current conservation and a dual of the current can be naturally
defined through $\epsilon^{ab}$. However, there is no analogue of
$\epsilon^{ab}$ in superspace. We circumvent this difficulty by
formulating the theory in a first order formulation with an
auxiliary spinor superfield. This allows us to define the dual
metric and the antisymmetric tensor fields in a manner
completely parallel to the bosonic string. We have presented the
explicit forms of the supercharges for both the original as well
as the dual theory. It will be of interest to study the question
of duality for more general backgrounds as in ref. 12.

We acknowledge collaboration with T. Turgut in the initial stage of
this work. We would like to thank Gabriele Veneziano for useful
discussions. This work was completed while one of us (A.D.) was a
visiting Scientist at the Institute of Physics. This work is partly
supported by U. S. Department of Energy Grant No. DE-FG-02-91ER40685.

\vfill
\eject

\centerline {\bf { References }}

\item{1.} K. Kikkawa, M. Yamasaki, Phys. Lett. {\bf B149}
(1984) 357; N. Sakai and I. Senda, Prog. Th. Phys. {\bf
75} (1986) 692. For a recent review see "Space time duality in
string theory", J. H. Schwarz in Elementary Particles and the
Universe, Essays in Honor of Murray Gell Mann, Ed J. H. Schwarz,
Cambridge University Press 1991.

\item{2.}  V.P. Nair, A. Shapere, A. Strominger and F. Wilczek, Nucl.
Phys. {\bf B287} (1987) 402; A. Shapere and F. Wilczek, Nucl. Phys.
{\bf B320} (1989) 669; T.H. Buscher, Phys. Lett. {\bf B194} (1987) 59
and {\bf 201} (1988) 466; A. Giveon, E. Ravinovici and G. Veneziano,
Nucl. Phys. {\bf B322} (1989) 169.

\item{3.}  S. Celcotti, S. Ferrara and L. Girardello, Nucl. Phys. {\bf
B308} (1988) 436; P. Ginsparg and C. Vafa, Nucl. Phys. {\bf B289} (1987)
414; M. Dine, P. Huet and N. Seiberg, Nucl. Phys. {\bf B322} (1989) 301;
J. Lauer, J. Mas and H.P. Nilles, Phys. Lett. {\bf B226} (1989) 251; W.
Lerche, D. L\"ust and N. Warner, Phys. Lett. {\bf B231} (1989) 417; B.R.
Greene and M.R. Plesser, Nucl. Phys. {\bf B338} (1990) 15; M.J. Duff,
Nucl. Phys. {\bf B335} (1990) 610; J.M. Molew and B. Ovrut, Phys. Rev.
{\bf D40} (1989) 1146; T. Banks, M. Dine, H. Dijrstra and W. Fischler,
Phys. Lett. {\bf B212} (1988) 45; A. A. Tseytlin Nucl. Phys. {\bf B350} (1991)
395; Phys. Rev. Lett. {\bf 66} (1991) 545.

\item{4.}  E. Alvarez and M. Osorio, Phys. Rev. {\bf D401} (1989) 1150.

\item{5.}  R. Brandenberger and C. Vafa, Nucl. Phys. {\bf B316} (1988)
391; M. Mueller, Nucl. Phys. {\bf B377} (1990) 37;
A.A. Tseytlin, Mod. Phys. Lett. {\bf 6A} (1991) 1721; A.A. Tseytlin and
C. Vafa, Nucl. Phys. {\bf B372} (1992) 443.

\item{6.}  G. Veneziano, Phys. Lett. {\bf B205} (1991) 287.

\item{7.}  A. Giveon, Mod. Phys. Lett. {\bf A6} (1991) 2843;
E. Smith and J. Polchinski, Preprint UTTG-07-91; M.
Rocek and E. Verlinde, Nucl. Phys. {\bf B373} (1992) 630; A.
Giveon and M. Rocek, Nucl. Phys. {\bf B380} (1992) 128.

\item{8.}  K.S. Narain, Phys. Lett. {\bf B169} (1986) 61; K.S. Narain,
H. Sarmadi and E. Witten, Nucl. Phys. {\bf B279} (1987) 309.

\item{9.}  K.A. Meissner and G. Veneziano, Phys. Lett. {\bf B267} (1991)
33; K.A. Meissner and G. Veneziano, Mod. Phys. Lett {\bf A6} (1991) 3397;
M. Gasperini, J. Maharana and G. Veneziano, Phys. Lett. {\bf B272} (1991)
167; and Phys. Lett. {\bf B296} (1992) 51; M. Gasperini and G. Veneziano,
Phys. Lett. {\bf B277} (1992) 256;
J. Panvel, Preprint LTH 282 DAMTP, University of Liverpool.

\item{10.}  A. Sen, Phys. Lett. {\bf B272} (1992) 34; F. Hassan and A. Sen,
Nucl. Phys. {\bf B375} (1992) 103;  S.P. Khastgir and A. Kumar,
Mod. Phys. Lett. {\bf 6A} (1991) 3365.

\item{11.} J. Maharana, Phys. Lett. {\bf B296 } (1992), 65.

\item{12.} J. Maharana and J. H. Schwarz, Nucl. Phys. {\bf B390}
(1993), 3.

\item{13.}  A .Sen and J. H. Schwarz, Phys. Lett. {\bf B312} (1993)
105, and preprint Duality Symmetric Action hep-th/ $9304154$.

\item{14.}  S. Pratik Khastgir and J. Maharana Phys. Lett. {\bf B301}
(1993) 191; Nucl. Phys. {\bf B406} (1993) 145; Classical and Quantum
Gravity {\bf10} (1993) 1731.

\item{15.}  E. B. Kiritsis, Proc. High Energy Conf. Marseille July 1993.
and references therein.

\item{16.}  C. Kounnas, Proc. High Energy Conf. Marseille July 1993
and references therein.

\item{17.}  For a recent review see L. Alvarez-Gaume CERN
Preprint CERN-TH 7036/93 and references therein.

\item{18.} Duality transformation in superstring theories has been
considered by W. Siegel, Phys. Rev. {\bf D48} (1993) 2826.

\bye